\documentclass[a4paper,11pt]{article}
\usepackage{subcaption}
\usepackage{epsfig, graphicx, subfloat, float}
\usepackage{latexsym}
\usepackage{cite}
\usepackage{amssymb}
\usepackage{latexsym}
\usepackage{mathrsfs}
\usepackage{amsmath}
\usepackage{color}
\usepackage[usenames]{xcolor}
\graphicspath{{Image/}}
\author{H. Mohseni Sadjadi \footnote{mohsenisad@ut.ac.ir}
\\ {\small Department of Physics, University of Tehran,}
\\ {\small P. O. B. 14395-547, Tehran 14399-55961, Iran}}
\title {Early dark energy and scalarization in a scalar-tensor model}
\begin{document}
\maketitle

\begin{abstract}
We present a model in which the Gauss-Bonnet invariant holds the quintessence at a fixed point, respecting an initial $Z_2$ symmetry in the radiation-dominated era. This results in an early dark energy that becomes significant around the matter-radiation equality era. However, due to $Z_2$ symmetry breaking, scalarization occurs, leading to a rapid reduction in the early dark energy density. The model then quickly behaves like the $\Lambda$CDM model. This scenario may also explain the reduction of sound horizon and aligns with the assumption that the gravitational wave speed is infinitesimally close to the speed of light.
\end{abstract}

\section{Introduction}
Introducing a cosmological constant ($\Lambda$), into the Einstein equation provides a straightforward explanation for the observed acceleration in the later stages of our universe\cite{acc1,acc2,acc3,acc4,acc5,acc6}. Although the $\Lambda$ Cold Dark Matter model ($\Lambda$CDM) yields promising results, it encounters issues such as the cosmological constant problem \cite{wein} and the coincidence problem \cite{coin1,coin2,coin3,coin4}. Consequently, one may consider alternative dark energy candidates, such as a slowly varying scalar field\cite{quint1,quint3,quint4,quint5,quint6,quint7,quint8,quint9,quint10,quint11,quint12,quint13}, to elucidate this late-time acceleration. A nearly constant scalar field with a potential $V(\phi)$ replicates the role of the cosmological constant in the background evolution, with energy density $\rho_\Lambda=V(\phi)$. However, the energy density of a component whose equation of state (EoS) parameter satisfies $w<-\frac{1}{3}$ dilutes more slowly than matter and radiation. Therefore, as in the present era matter and dark energy densities have comparable magnitudes, the relative dark energy density should have been negligible in the early Universe unless another mechanism, beyond the usual redshift, diminished the significant early dark energy density.

Among dark energy models those that modify the early Universe may provide a viable solution to the Hubble tension \cite{Hub0} (i.e. the disparity between the Hubble parameter calculated by inverse distance ladder and low redshift measurements \cite{Hub1,Hub2,Hub3,Hub4}). Early dark energy (EDE), which transiently becomes significant around matter-radiation equality, has recently been utilized to address the Hubble tension \cite{EDE1,EDE2,EDE3,EDE4,EDE5,EDE6,EDE7,EDE8,EDE9,EDE10,EDE11}.
While EDE has demonstrated potential in addressing the Hubble tension, this is not true for all EDE models \cite{EDE5}.
Models incorporating a dynamical scalar field as EDE have also been proposed to resolve the Hubble tension, as seen in \cite{EDE2,EDE3,EDE4}. The current Hubble parameter satisfies $H_0\propto \frac{\theta_*}{r_s}$, where $\theta_*$ is the angular size on the last scattering surface, and $r_s$ is the sound horizon of acoustic waves in photon baryon fluid. Since $\theta_*$ is solely and precisely determined from the first cosmic microwave background peak, an increase in $H_0$, can be explained by a decrease of $r_s$. An EDE, lowering the sound horizon by $\sim 7\%$, may alleviate the Hubble tension \cite{EDE6,hun,ten1,ten2}.

Despite EDE garnering significant attention in recent years, the exploration of EDE predates the emergence of the Hubble tension problem, as evidenced by studies such as \cite{ed1,ed2,ed3,ed4,ed5,ed6}. An EDE model with an equation of state (EoS) parameter $w=-1$ was employed in \cite{ed4} to investigate the absorption of Cosmic Microwave Background (CMB) photons by the $21cm$ hyperfine transition of neutral hydrogen, a phenomenon reported by the Experiment to Detect the Global Epoch of Reionization Signature (EDGES) collaboration. EDE can lead to an earlier decoupling of gas temperature from radiation temperature. In \cite{ed5}, using a phenomenological parametrization of dark energy density across different eras, an upper bound for relative EDE density was suggested, with $\Omega_d<0.06$. In another study \cite{ed6}, employing a parameterized energy model, the authors found an upper limit of $\Omega_d<2.6\%$ during the radiation-dominated era and $\Omega_d<1.5\%$ within the redshift range $z\in (100,1000)$. It is essential to note that these papers distinguish between EDE and late dark energy based on a phenomenological parameterized approach, with both contributing separately to the total density.

An alternative approach to discussing cosmic positive acceleration involves modifying Einstein's theory of gravity by introducing geometric terms, such as the Gauss-Bonnet (GB) invariant, into the action. In four dimensions, the GB invariant only contributes a surface term and does not alter the Einstein equation. However, when coupled with exotic fields like quintessence, it induces significant and intriguing effects on cosmic evolution, particularly late-time acceleration and super-acceleration\cite{GB0,GB1,GB2,GB3,GB33,GB4,GB5,GB6,GB7,GB8,GB9,GB10,GB11}. Recently, constraints placed on the gravitational wave speed have raised doubts about the direct influence of the GB model on cosmic evolution at low redshifts \cite{ob1,ob2,ob3,ob4,doub,doub1}. The scalarization of black holes and neutron stars, within the context of the scalar-GB model, has also garnered considerable attention. In a scalar-tensor model comprising a scalar field coupled to the GB invariant, neutron stars, and black holes could exhibit scalar fields and behave differently than the standard general relativity (GR). The scalarized solution may be triggered by a tachyonic instability \cite{sc1,sc2,sc3,sc4}.

By relating the three seemingly unrelated aforementioned topics, i.e. EDE, scalarization, and the gravitational wave speed, this paper presents a model where the quintessence-GB coupling establishes conditions for an initial stable fixed-point solution, representing an EDE. The relative density of this EDE becomes significant around the matter-radiation equality era and then rapidly decreases to align with the usual $\Lambda$CDM model. This reduction is attributed to a tachyonic instability induced by radiation and matter dilution, triggering quintessence evolution and its emergence through a scalarization-like mechanism. We demonstrate that constraints on the gravitational wave speed associated with the GB coupling align with this scenario.
Note that in our study, we assume that the gravitational wave speed is very close to the light speed and lies in the domain reported in \cite{ob4}.

The paper is structured as follows: Section 2 provides a detailed presentation of the model, emphasizing how the dilution of matter and radiation density initiates quintessence evolution through its coupling with the GB term, resulting in the aforementioned scenario. We explore stability conditions and investigate the impact of gravitational wave speed constraints on quintessence activation. To show how the model works, we illustrate our results with a specific example. In Section 3, we provide concluding remarks.

We use units $\hbar=c=1$.

\section{Early dark energy and scalarization in $Z_2$ symmetric Gauss-Bonnet model }
We consider the following action describing a quintessence field $\phi$ coupled to the GB term \cite{GB5}:
\begin{equation}\label{1}
S=\int d^4x \sqrt{-g}\left(\frac{M_P^2 R}{2}-\frac{1}{2}g^{\mu \nu}\partial_\mu \phi \partial_\nu \phi-V-
\frac{1}{2}f\mathcal{G}\right)+S_m.
\end{equation}
$M_P=2.4\times 10^{18} GeV$ is the reduced Planck mass. The quintessence, $\phi$, is coupled to the GB term through the coupling function $f:=f(\phi)$. The quintessence
potential is $V:=V(\phi)$. The GB invariant is given by
\begin{equation}\label{2}
\mathcal{G}=R_{\mu \nu \rho \sigma}R^{\mu \nu \rho \sigma}-4R_{\mu \nu}R^{\mu \nu}+R^2.
\end{equation}
We consider a spatially flat Friedmann-Lema\^{\i}tre-Robertson-Walker (FLRW) space-time,
\begin{equation}\label{3}
ds^2=-dt^2+a^2(t)(dx^2+dy^2+dz^2),
\end{equation}
filled with the quintessence, cold dark matter, baryonic matter,
and radiation. Friedmann equations read
\begin{eqnarray}\label{4}
&&3M_P^2H^2=\rho_d+\rho_r+\rho_m \nonumber \\
&&2M_P^2\dot{H}=-\rho_d-P_d-\rho_m-\frac{4}{3}\rho_r.
\end{eqnarray}
$H$ is the Hubble parameter, which in terms of the scale factor $a$, is given by $H=\frac{\dot{a}}{a}$. $\rho_m$ and $\rho_r$ are the sum of baryonic and cold dark matter, and radiation energy densities respectively. These barotropic fluids satisfy the continuity equations
\begin{eqnarray}\label{5}
\dot{\rho_m}+3H\rho_m=0\nonumber \\
\dot{\rho_r}+4H\rho_r=0.
\end{eqnarray}
The effective dark energy density and pressure are given by
\begin{equation}\label{6}
\rho_d=\frac{1}{2}\dot{\phi}^2+V+12H^3\dot{f},
\end{equation}
and
\begin{equation}\label{7}
P_d=\frac{1}{2}\dot{\phi}^2-V-8H^3\dot{f}-4H^2\ddot{f}-8H\dot{H}\dot{f},
\end{equation}
respectively. For the effective dark energy, we have the following continuity equation:
\begin{equation}\label{8}
\dot{\rho_d}+3H(P_d+\rho_d)=0,
\end{equation}
which is equivalent to the following equation of motion
\begin{equation}\label{9}
\ddot{\phi}+3H\dot{\phi}+V^{eff.}_{,\phi}=0,
\end{equation}
where we have defined \cite{sad}
\begin{equation}\label{10}
V^{eff.}_{,\phi}:=V_{,\phi}+12H^2(H^2+\dot{H})f_{,\phi}.
\end{equation}
The subscript ${,\phi}$ denotes derivatives with respect to $\phi$.
It is evident from (\ref{9}) that the quintessence evolution is influenced by the GB invariant. We use this to introduce a scalarization scenario in which the system acquires a nontrivial quintessence solution through a tachyonic instability.

We construct our model such that the action (\ref{1}) has a $Z_2$ symmetry (i.e. is invariant under $\phi\to -\phi$). So we take $V$ and  $f$ as even functions of $\phi$ leading to $V_{,\phi}(\phi=0)=f_{,\phi}(\phi=0)=0$. As a consequence $V^{eff.}_{,\phi}(\phi=0)=0$.  Therefore $\phi=0$ becomes a trivial solution to (\ref{9}), respecting the $Z_2$ symmetry. From (\ref{4}), we find out that for this solution the Friedmann equations (\ref{4}) reduce to ordinary (non-modified) ones:
\begin{eqnarray}\label{11}
&&3M_P^2H^2=V(0)+\rho_r+\rho_m \nonumber \\
&&2M_P^2\dot{H}=-\rho_m-\frac{4}{3}\rho_r.
\end{eqnarray}
The dark energy density is given by $\rho_d=V(0)$, playing the role of a cosmological constant. We aim to introduce a cosmological scalarization via a tachyonic instability that causes the model to attain a nontrivial (non-constant) scalar solution. This can be achieved by taking the initial restoring $Z_2$ symmetry solution as an initial stable fixed point which becomes unstable due to the evolution of the Universe. This is somehow similar to the symmetron dark energy model, where the quintessence evolution is triggered by matter dilution \cite{sym}.

As $V^{eff.}_{,\phi}(0)=0$, the $\phi=0$ solution is stable (unstable) when $V^{eff.}_{,\phi\phi}(0)>0(<0)$, corresponding to the minimum (maximum) of the effective potential. Hence the quintessence stability at $\phi=0$ depends on the sign of $V^{eff.}_{,\phi\phi}(0)$ given by
\begin{equation}\label{12}
\begin{split}
M_{eff.}^2&\equiv V^{eff.}_{,\phi\phi}(0)=-\frac{2}{3M_P^4}(\rho_r+\rho_m-2V(0))(\rho_r+\rho_m+V(0))f_{,\phi \phi}(0)\\
&+V_{,\phi \phi}(0).
\end{split}
\end{equation}
We define the critical scale factor, denoted as $a_c$, by $M_{eff.}^2(a_c)=0$ and $\frac{dM_{eff.}^2}{da}(a_c)<0$. For $\phi=0$ to be a stable solution before $a_c$, it is necessary to have $V^{eff.}_{,\phi\phi}(0)>0$ for $a<a_c$, implying $f_{,\phi \phi}(0)<0$.  $M_{eff.}^2$ can be considered as the effective mass squared of small fluctuations around $\phi=0$ for $a\leq a_c$. By choosing
$V_{,\phi \phi}(0)<0$, we may have  $M_{eff.}^2\leq 0$ for  $a\geq a_c$, and the quintessence becomes tachyonic. In this situation, $\phi=0$ becomes an unstable point at the local maximum of the effective potential. Through a small fluctuation, the quintessence rolls down its effective potential, and its evolution begins and the system gets a non-trivial solution that no longer respects the $Z_2$ symmetry. The emergence of the scalar field also modifies the standard Friedmann equations from (\ref{11}) to (\ref{4}), where the GB term directly influences the expansion of the Universe. Note that to determine the behavior of the scalar field for $a>a_c$, we have to solve one of the Friedmann equations (\ref{4}), and the equations of motions (\ref{5}), (\ref{9}).

As we have assumed $V_{,\phi \phi}(0)<0$, the quintessence evolution along the effective potential decreases $V$, leading to a reduction in early dark energy (EDE) density. Indeed, in a model with a non-negligible EDE, as the dark energy density dilutes more slowly than ordinary and dark matter, it is crucial to introduce a process other than the usual redshift to reduce the dark energy density. To clarify this statement, we make use of the relation
\begin{equation}\label{13}
\frac{\rho_d}{\rho_{d0}}=\frac{\Omega_d}{1-\Omega_d}\left(\frac{\Omega_{m0}a^{-3}+
\Omega_{r0}a^{-4}}{\Omega_{d0}}\right),
\end{equation}
where $\Omega_i=\frac{\rho_i}{3M_P^2 H^2}$ represents the relative density at an arbitrary scale factor $"a"$, and the subscript $"0"$ denotes the present time specified by $a=1$. We can use astrophysical data \cite{Planck_data},
to set $\Omega_{m0}\simeq 0.32, \Omega_{r0}\simeq 8.4\times 10^{-5}$.  for DE whose effective EoS parameter satisfies $w_d^{eff.}<-\frac{1}{3}$, the redshift dilution is derived as $\frac{\rho_d}{\rho_{d0}}=a^{-3(1+w_d^{eff.})}< a^{-2}$.
This, for a non-negligible $\Omega_d(a)$ at high redshifts, is much smaller than what is given by (\ref{13}).  For example by taking $\Omega_d=0.01$ at a scale factor $a \lesssim 10^{-4}$ (before matter radiation equality), (\ref{13}) gives $\frac{\rho_d}{\rho_{d0}}\gtrsim 2.02\times 10^{10}$ which is at least two order of magnitudes larger than  $\frac{\rho_d}{\rho_{d0}}<10^8$ obtained from the redshift. In this example, the usual redshift is consistent with (\ref{13}) provided that $\Omega_d (a \lesssim 10^{-4})<0.004$. In our model, the scalarization occurs after $a=a_c$. For $a\leq a_c$ the dark energy density is given by $V(\phi=0)$ which is a cosmological constant (with $w_d=-1$). The fractional density in the absence of the scalarization may be derived by setting $\frac{\rho_d}{\rho{d_0}}=1$ in (\ref{13}), which for the above numeric example gives $\Omega_d(a\lesssim 10^{-4})\lesssim 5\times 10^{-11}$ which is negligibly small.

Hence, to reconcile non-negligible EDE and late Dark energy, an additional mechanism, beyond the redshift dilution, is required to reduce the significant EDE density by several orders of magnitude. In our scenario, this reduction is achieved through the activation of the quintessence via the discussed scalarization. For this purpose, {\it {we choose the potential as a decreasing function of $\phi^2$}}. Unlike models where quintessence activation drives cosmic acceleration, in our case, its activation leads to a reduction in dark energy density. Such a formalism is necessary for models considering a temporarily significant EDE, e.g. around the matter-radiation equality.

It is important to note that, owing to the presence of the friction term, the evolution of the scalar field does not begin immediately after $a=a_c$. Therefore, for the desired scenario of having significant dark energy and its subsequent reduction around the matter-radiation equality era, and also a very tiny coupling to the Gauss-Bonnet invariant which is consistent with the gravitational wave speed (\ref{17}) \cite{sad}, the critical scale factor $a_c$ is favored to be within the radiation-dominated era, although the capacity of this model to resolve the Hubble tension in agreement with cosmological data such as large-scale structures etc. should be verified with a detailed statistical analysis.

The critical scale factor, $a_c$, is obtained by solving the equation
\begin{eqnarray}\label{14}
&&V_{,\phi \phi}(0)=\frac{2}{3M_P^4}(\rho_{r0}a_c^{-4}+\rho_{m0}a_c^{-3}-2V(0))(\rho_{r0}a_c^{-4}+\rho_{m0}a_c^{-3}\nonumber \\
&&+V(0))f_{,\phi \phi}(0),
\end{eqnarray}
where $\rho_{i0}$ is the energy density at the present time ($a_0=1$). As evident from (\ref{14}), smaller values of $a_c$  result in tinier values of $f_{,\phi \phi}$, which, as will be seen, is in favor of the GWS constraint.
By considering the perturbed space-time
\begin{equation}\label{15}
ds^2=-dt^2+a^2(t)(t_{ij}+\delta_{ij})dx^idx^j,
\end{equation}
 one can obtain a second-order action for divergenceless  and traceless $t_{ij}$, and obtain GWS. For the model (\ref{1}) which is a special case of Horndeski's theories, the gravitational wave speed (GWS) is obtained as \cite{speed1,speed2}
\begin{equation}\label{16}
c_{T}^2=\frac{4f_{,\phi\phi}\dot{\phi}^2+4f_{,\phi}\ddot{\phi}-M_P^2}{4Hf_{,\phi}\dot{\phi}-M_P^2}=
\frac{4\ddot{f}-M_P^2}{4H\dot{f}-M_P^2}.
\end{equation}
According to \cite{ob4}, this speed is constrained at the late time for the redshift $z<0.009$ as:
\begin{equation}\label{17}
-3\times 10^{-15}\leq \frac{c_{T}}{c}-1 \leq 7\times 10^{-16},
 \end{equation}
 where $c$ is the light speed. To satisfy (\ref{17}), generally we must take  $4H\dot{f}\ll M_P^2$ and  $\ddot{f}\ll M_P^2$, which implies that a nearly constant $\phi$ with infinitesimal couplings to the GB term are required. For a quadratic $f$ a tiny coupling favors that $a_c$ is in the radiation era. A nearly constant quintessence at the late time is realized if the potential becomes nearly flat: $V(\phi)\simeq \Lambda$, $V_{,\phi}^2\ll H^2 V$, $V_{,\phi \phi}\ll H^2$. In such a situation the quintessence slowly rolls $\frac{1}{2}\dot{\phi}^2\ll V(\phi)$.  Note that $4H\dot{f}\ll M_P^2$ and $\ddot{f}\ll M_P^2$ results in that the GB term has no direct influence in cosmic evolution in the late time. This can be seen from the Friedmann equations, rewritten as \cite{ts}
\begin{eqnarray}\label{sh}
&&3(M_P^2-4H\dot{f})H^2=\frac{1}{2}\dot{\phi}^2+V+\rho_r+\rho_m \nonumber \\
&&2(M_P^2-4H\dot{f})\dot{H}=-\dot{\phi}^2-(M_P^2-4H\dot{f})(c_T^2-1)\nonumber \\
&&-\rho_m-\frac{4}{3}\rho_r.
\end{eqnarray}

The smallness of the quintessence-GB coupling also helps the stability of the model, against ghost and Laplacian instabilities caused by scalar and tensor perturbations, which requires that the following inequalities hold \cite{ts}:
\begin{eqnarray}\label{18}
&&q_T:=M_P^2-4f_{,\phi}\dot{\phi}H>0 \nonumber \\
&& q_s:=2(q_T+24H^4f_{,\phi}^2)>0\nonumber \\
&&c_s^2=\frac{2q_T-16H^4f_{,\phi}^2(2+6w+c_T^2)}{q_s}>0.
\end{eqnarray}
$w=-1-\frac{2}{3}\frac{\dot{H}}{H^2}$ is the Universe effective equation of state.

A recent motivation to consider the EDE is to alleviate the Hubble tension. This can done by an EDE component which becomes significant around the matter-radiation equality era and then decreases quickly through a mechanism other than the redshift, similar to what happens in our proposal. This decreases the sound horizon and results in a larger value for the present Hubble parameter through the relation $H_0\propto \frac{\theta_*}{r_s}$, where $\theta_*$ is the angular size on the last scattering surface determined from the first cosmic microwave background peak and accurately measured from the CMB observations by missions like Planck. So the comoving sound horizon at recombination, $r_s$, is pivotal in understanding the Hubble tension because it directly impacts the inferred value of the Hubble constant. At the last scattering, $r_s$, is derived as \cite{ed6},\cite{EDE4}
\begin{eqnarray}\label{19}
&&r_s=\int_{z_{*}}^{\infty}\frac{c_s(z)}{H(z)}dz =\int_{z_{*}}^{\infty}\frac{c(z)}{\sqrt{\frac{1}{3M_P^2}\sum_i \rho_i}}dz\nonumber \\
&&=\frac{1}{H_0}\int_{z_{*}}^{\infty}dz\frac{c(z)}{\sqrt{\Omega_{r0}(1+z)^{4}+\Omega_{m0}(1+z)^{3}+\frac{\rho_d}{3M_P^2H_0^2}
}},
\end{eqnarray}
$\Omega_m=\Omega_{dm0}+\Omega_{b0}$, where $\Omega_{b0}$ and $\Omega_{dm0}$ are relative densities of dark matter and baryonic matter at $a=1$, respectively. $z_{*}$ is the redshift of the last scattering, and $c_s(z)$ is the sound speed in the baryon-photon fluid, given by \cite{sound}:
\begin{equation}\label{20}
c_s(z)=\frac{1}{\sqrt{3}}\left( \frac{3}{4}\frac{\Omega_{b0}}{\Omega_{r0}}\frac{1}{1+z}+1\right)^{-\frac{1}{2}}
\end{equation}
Adding an EDE increases $H(z)$, and consequently decreases $r_s$. This may be promising to resolve the Hubble tension \cite{hun}. If the EDE density increases effectively the energy density as $\sum \rho\to (1+\gamma)^2\sum\rho$ in (\ref{19}), then $H\to (1+\gamma)H$, and $r_s$ decreases by $\frac{100\gamma}{1+\gamma}\%$, e.g for $\gamma=0.075$, $r_s$ decreases by $\simeq 7\%$ \cite{hun}.

We continue our study with a specific example to illustrate how the model works. We choose the potential:
\begin{equation}\label{21}
V=V_0\exp(-\frac{1}{2}\mu^2\phi^2)+\Lambda,
\end{equation}
which is a decreasing function of $\phi^2$, consisting of a constant $\Lambda$ playing the role of a cosmological constant energy density at the late time, and a steep part which decreases rapidly for $\mu^2\phi^2\gtrsim 1$.
We adopt a quadratic coupling:
\begin{equation}\label{22}
f=-\frac{1}{2}\alpha^2\phi^2.
\end{equation}
We employ dimensionless parameters
\begin{eqnarray}
&&\hat{H}=\frac{H}{H^*},\,\, \hat{t}=H^*t,\, \hat{\mu}=M_P\mu,\, \hat{\alpha}=H^*\alpha,\,  \hat{\rho}=\frac{\rho}{M_P^2H^{*2}},\nonumber \\
&&\hat{V}_0= \frac{V_0}{M_P^2H^{*2}},\,\hat{\Lambda}= \frac{\Lambda}{M_P^2H^{*2}}
\end{eqnarray}
where $H^*$ is a mass scale. We set the initial conditions in the radiation-dominated epoch at $a=1/37500$ as
\begin{equation}\label{23}
\hat{\phi}=0,\,\, \hat{\phi}'=10^{-16},\,\, \hat{\rho}_m=10^{13},\,\, \hat{\rho}_r=10^{14}
\end{equation} a prime denotes derivative with respect to the dimensionless time $\hat{t}$. We select the following parameters
\begin{equation}\label{24}
\hat{\alpha}=10^{-7},\,\,\hat{\mu}=30,\,\, \hat{\Lambda}=0.04,\,\,  \hat{V}_0=4\times 10^8
\end{equation}
We have selected the parameters and initial conditions such that initially the relative EDE is insignificant $\Omega_d\simeq 10^{-6}$, where we have denoted the relative density by $\Omega_i\equiv \frac{\rho_i}{3M_P^2H^2}$. Note that the parameters and initial conditions satisfy (\ref{14}).
(\ref{14}) implies that $a_c$ has to be taken in the radiation dominated era to have a tiny $\hat{\alpha}$.  Note that the small value of $\hat{\alpha}$ along with the slowness of the quintessence rolling at the late time which is a consequence of the chosen potential, ensures compliance with the GWS constraint. $\hat{\Lambda}$ is chosen such that the model gives the correct dark energy density in our present era.

\begin{figure}[H]
\centering
\begin{subfigure}{.5\textwidth}
  \centering
  \includegraphics[width=.9\linewidth]{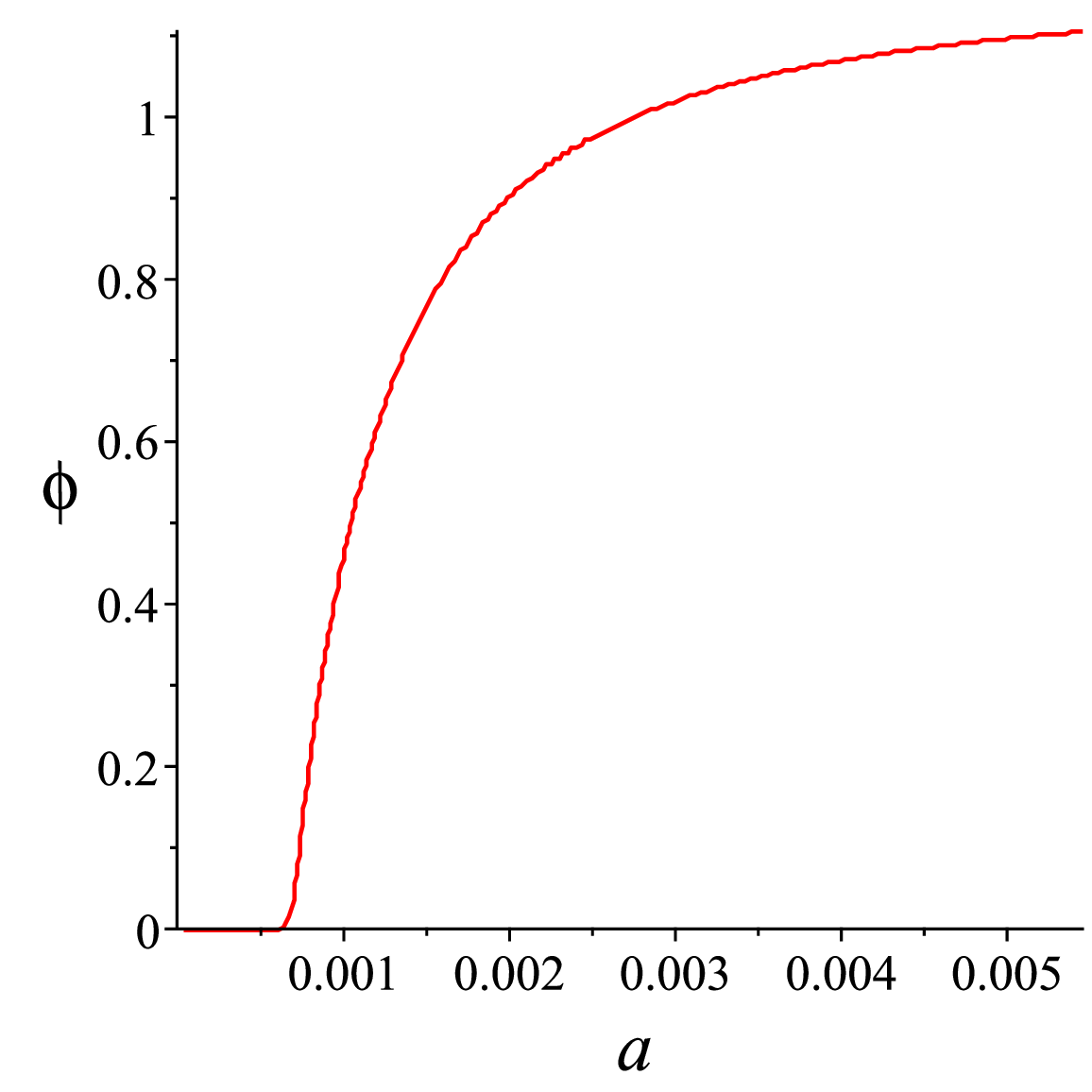}
  \caption{Beginning of the evolution}
  \label{fig:sub1}
\end{subfigure}%
\begin{subfigure}{.5\textwidth}
  \centering
  \includegraphics[width=.9\linewidth]{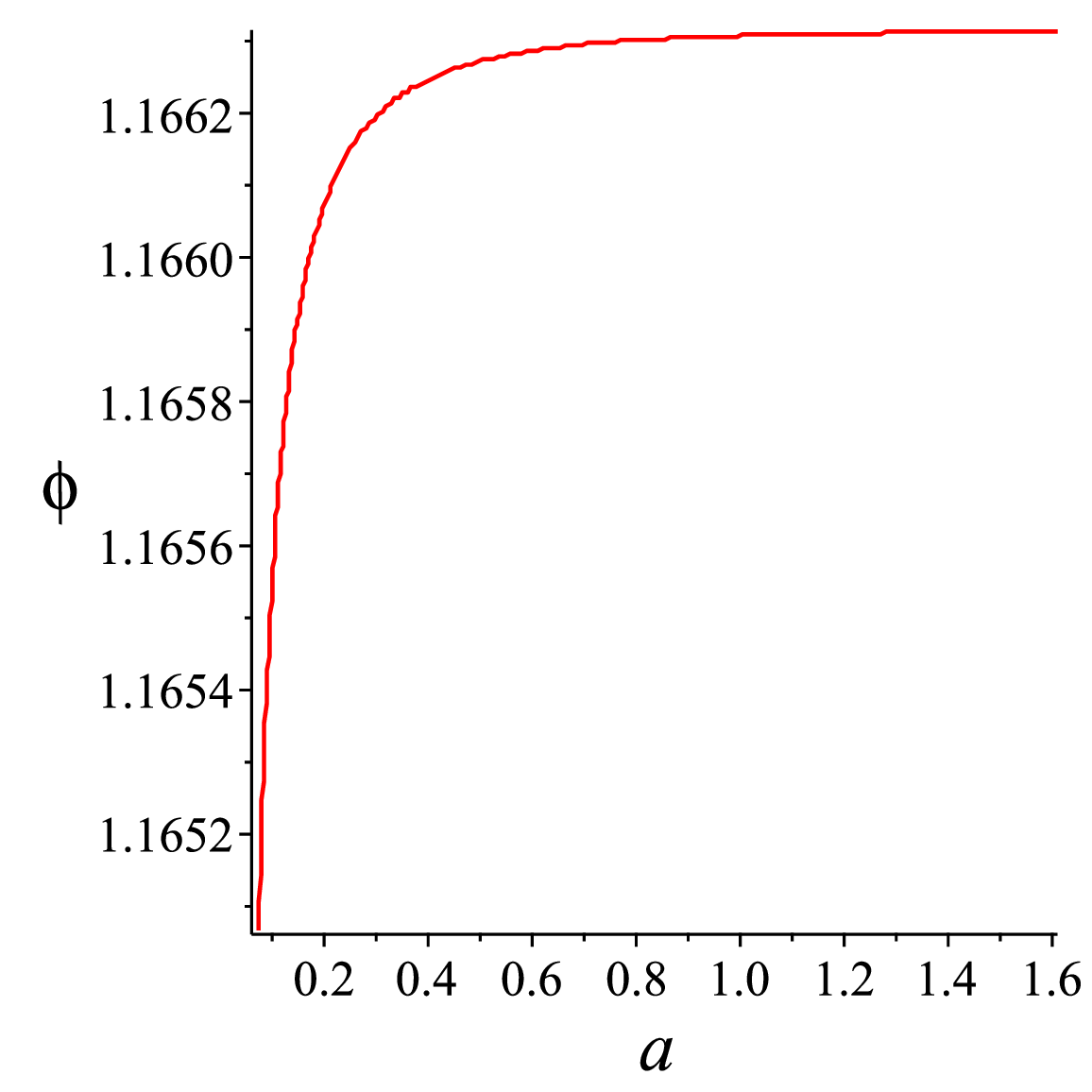}
  \caption{Late time evolution}
  \label{fig:sub2}
\end{subfigure}
\caption{Quintessence evolution versus the scale factor}
\label{fig1}
\end{figure}

Using the Friedmann equation (\ref{4}), the continuity equations (\ref{5}), and the equation of motion (\ref{9}), we can depict numerically the behavior of the system. We specify by $a=1$, $(z=0)$, our present time. The parameters are adjusted by confronting the results at $a=1$ with \cite{Planck_data}.
The evolution of the quintessence is depicted in Fig.(\ref{fig1})

Initially, we set $\phi$ to be equal to zero. This represents a trivial solution for the quintessence equation of motion. During the radiation era, the quintessence became tachyonic, but its evolution and emergence actually commenced around the time of matter-radiation equality, thanks to the friction term. At the stable point $\phi=0$, we have $V(0)=V_0+\Lambda$, which serves as an early cosmological constant energy density. When $\mu^2 \phi^2\gg \ln(\frac{V_0}{\Lambda})$, the potential becomes nearly flat: $V(\phi)\simeq \Lambda$, and it takes on the role of the actual cosmological constant. It is only between these two stages that the GB term may directly impact the Friedmann equations.

\begin{figure}[H]
\centering
\begin{subfigure}{.5\textwidth}
  \centering
  \includegraphics[width=.9\linewidth]{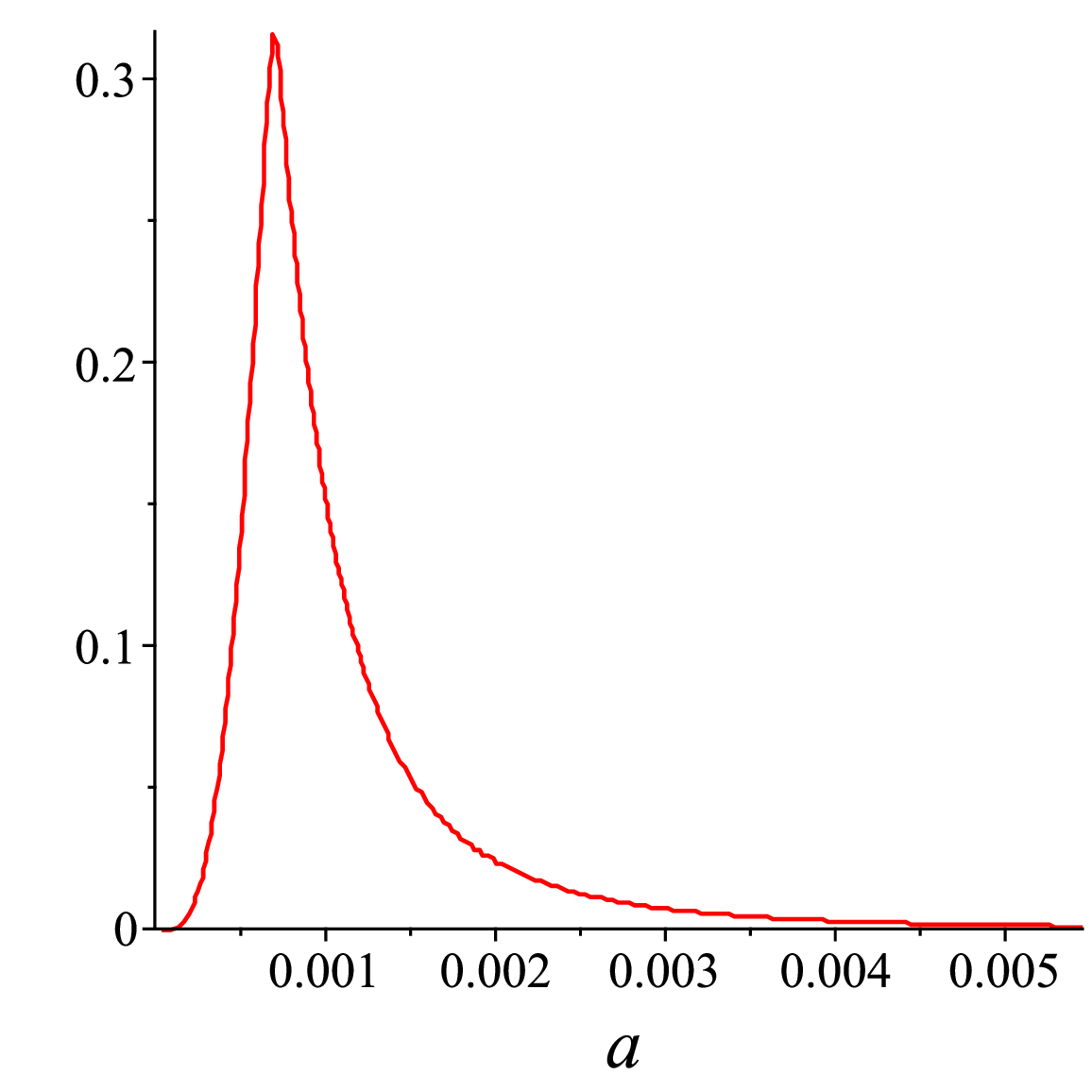}
  \caption{Early universe}
  \label{fig:sub1}
\end{subfigure}%
\begin{subfigure}{.5\textwidth}
  \centering
  \includegraphics[width=.9\linewidth]{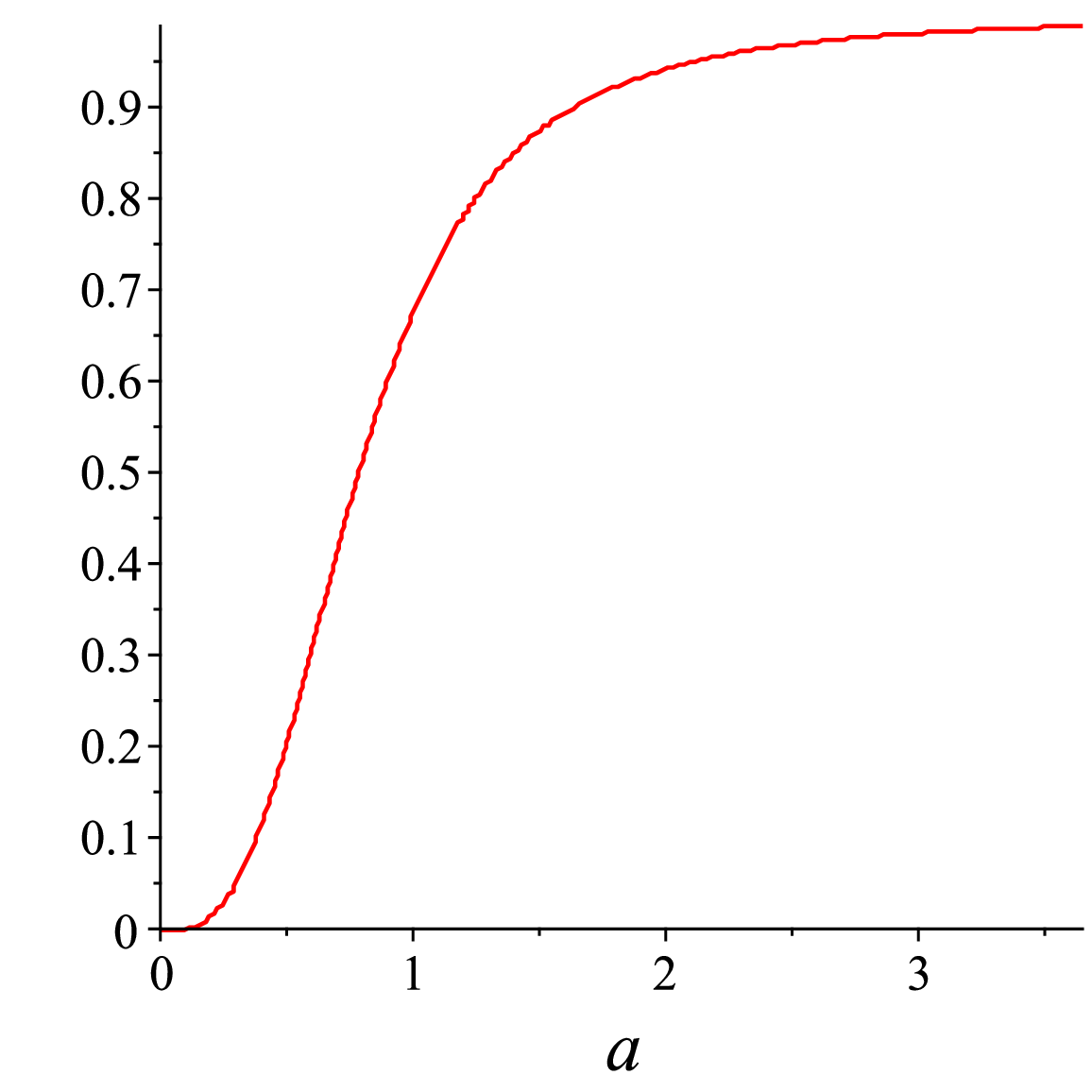}
  \caption{Late time}
  \label{fig:sub2}
\end{subfigure}
\caption{Relative dark energy density versus the scale factor}
\label{fig2}
\end{figure}

\begin{figure}[H]
	\centering
	\includegraphics[scale=0.3]{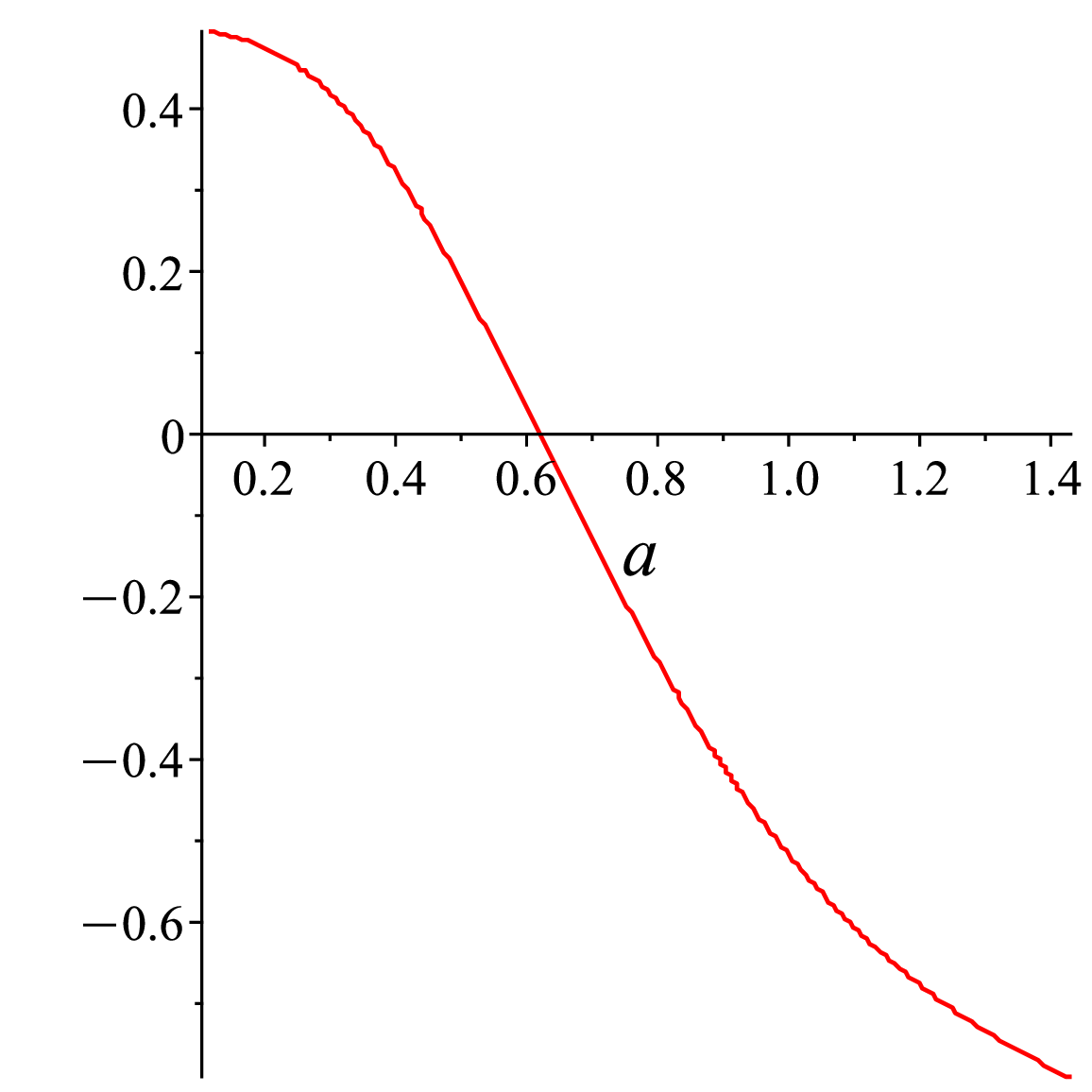}
	\caption{Deceleration parameter in terms of the scale factor}
	\label{dec}
\end{figure}

The relative density of dark energy is depicted in Fig.(\ref{fig2}), showing that the EDE is negligible for large redshifts in the radiation era. It increases until the quintessence becomes dynamic due to the tachyonic instability, and then due to the steep potential decreases quickly and behaves as the cosmological constant in the $\Lambda$CDM model. In this figure we have
$\Omega_d(a=1)=0.68$, which is compatible with the value reported in \cite{Planck_data}, on the base of $\Lambda$CDM model from Planck CMB power spectra in combination with CMB lensing (TT,TE,EE+lowE+lensing $68\%$ limit), as $\Omega_d= 0.6847 \pm 0.0073$.
The deceleration parameter, $q=-1-\frac{\dot{H}}{H^2}$ is depicted in Fig.(\ref{dec}), showing that the Universe entered the positive acceleration phase at $z\simeq 0.6$. In the present era, $q(a=1)=-0.521$ which corresponds to an effective EoS parameter $w(a=1)=-0.6807$ for the Universe.  This is compatible with \cite{Planck_data}, where the dark energy EoS parameter is reported as $w_d=-1.03\pm 0.03$ (note that in our era $w\simeq \Omega_d w_d$).

One can compute the sound horizon from (\ref{19}) numerically by using the set of equations (\ref{4}),(\ref{5}),(\ref{9}).
 Repeating the same computation while ignoring EDE one obtains $r_s^{\Lambda CDM}$. Taking the last scattering redshift at $z_{*}=1100$, for the above example we obtain $\frac{r_s^{\Lambda CDM}-r_s}{r_s^{\Lambda CDM}}\simeq 0.06$. This is compatible with the results reported by the SHOES team $H_0=72.1\pm 2.0 km/s/MPC$,
and reported in \cite{Planck_data} (TT,TE,EE+lowE+lensing+BAO $68\%$ limit )  $H_0=67.66\pm 0.42 km/s/MPC$.

\begin{figure}[H]
\centering
\begin{subfigure}{.5\textwidth}
  \centering
  \includegraphics[width=.9\linewidth]{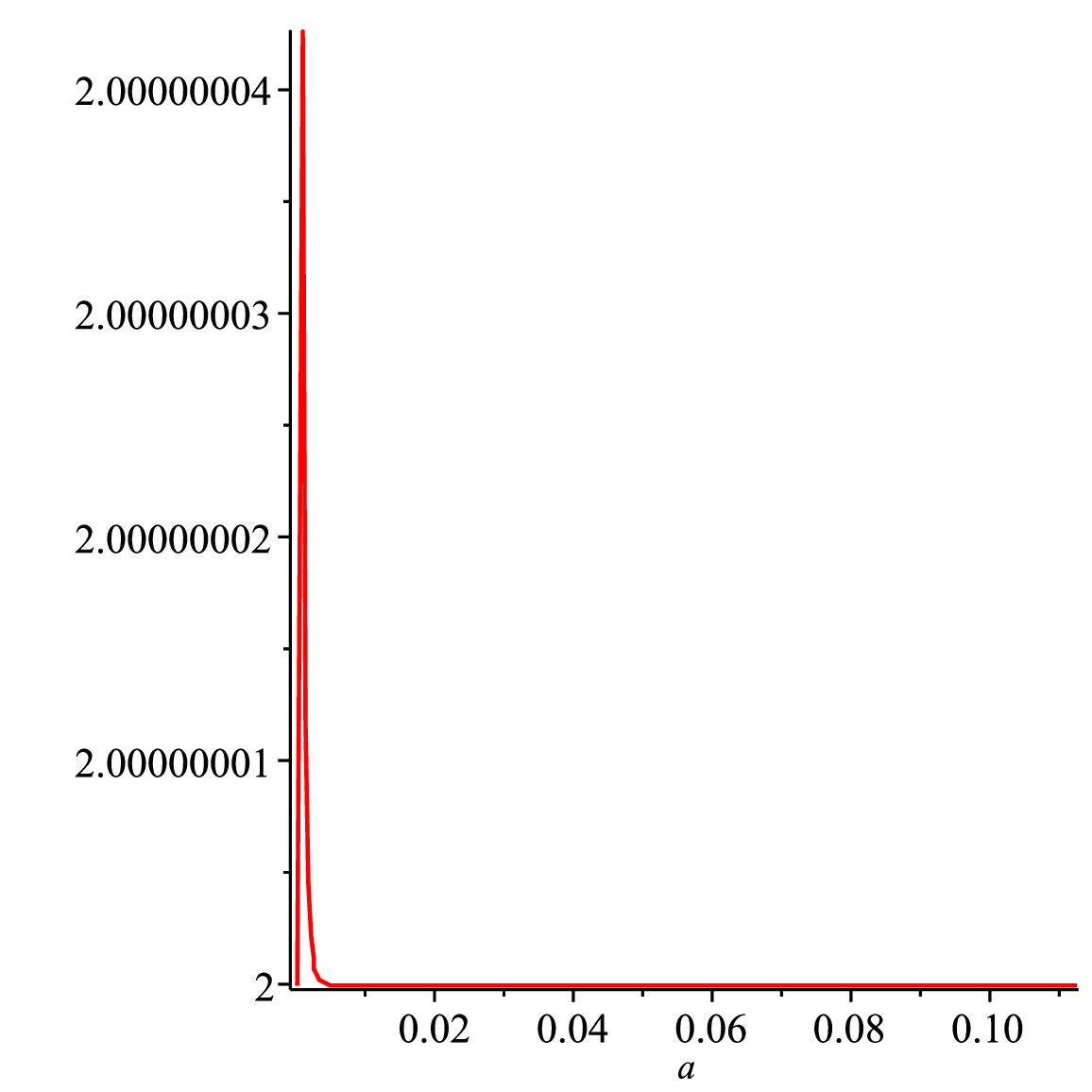}
  \caption{$q_s$ versus the scale factor}
  \label{fig:sub1}
\end{subfigure}%
\begin{subfigure}{.5\textwidth}
  \centering
  \includegraphics[width=.9\linewidth]{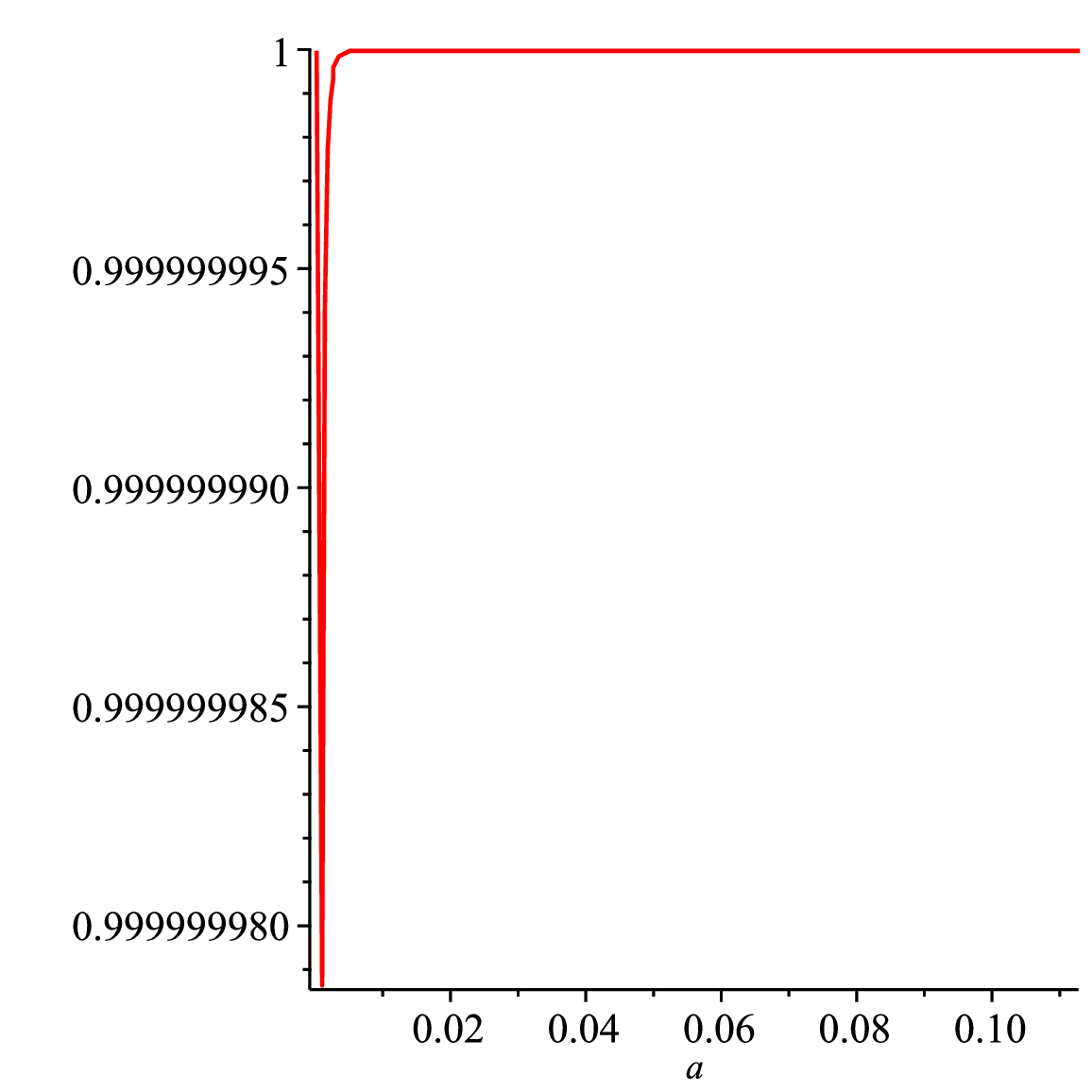}
  \caption{$q_T$ versus the scale factor}
  \label{fig:sub2}
\end{subfigure}
\caption{$q_T$ and $q_s$ versus the scale factor}
\label{fig3}
\end{figure}

\begin{figure}[H]
\centering
\begin{subfigure}{.5\textwidth}
  \centering
  \includegraphics[width=.9\linewidth]{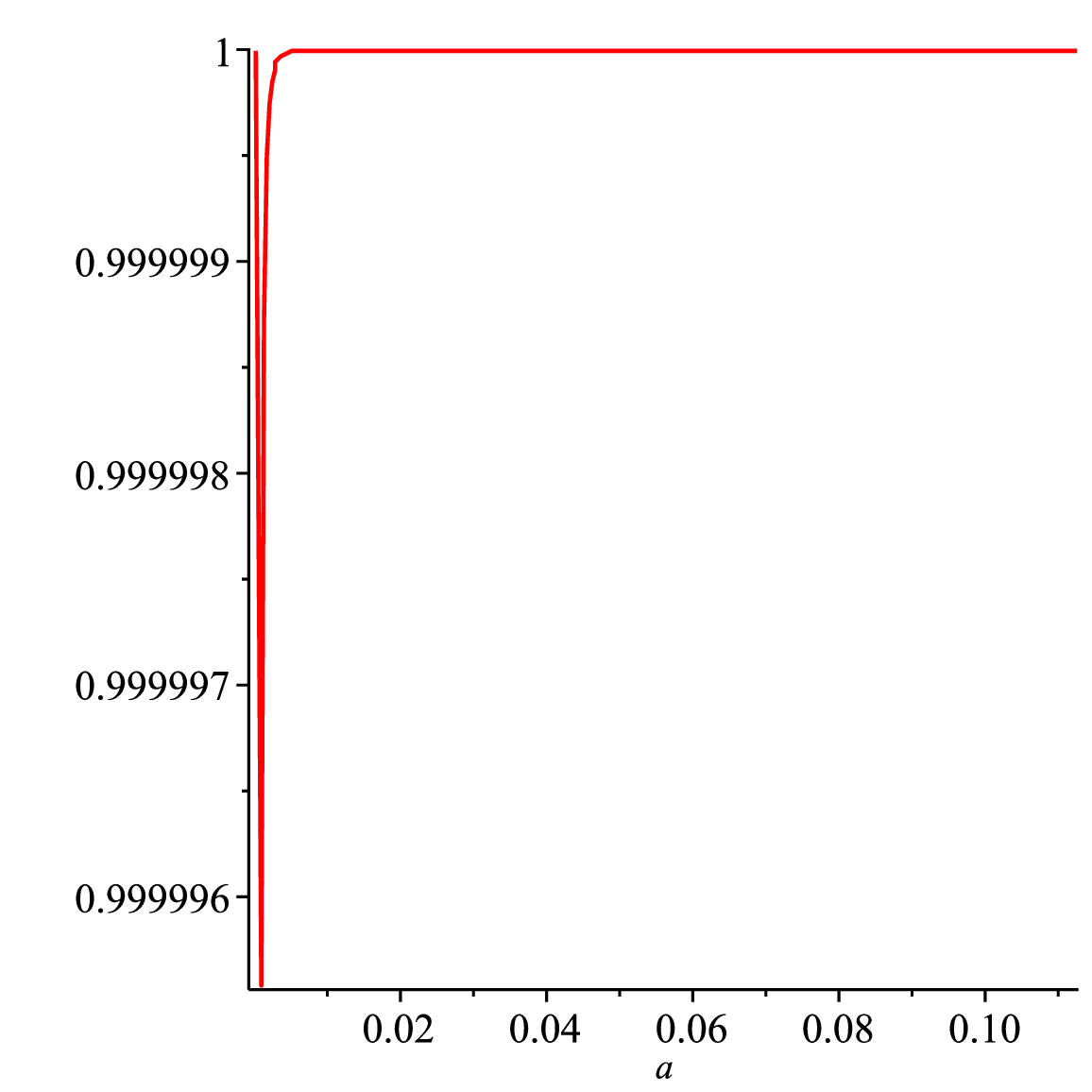}
  \caption{$c_s^2$ in terms of the scale factor}
  \label{fig:sub1}
\end{subfigure}%
\begin{subfigure}{.5\textwidth}
  \centering
  \includegraphics[width=.9\linewidth]{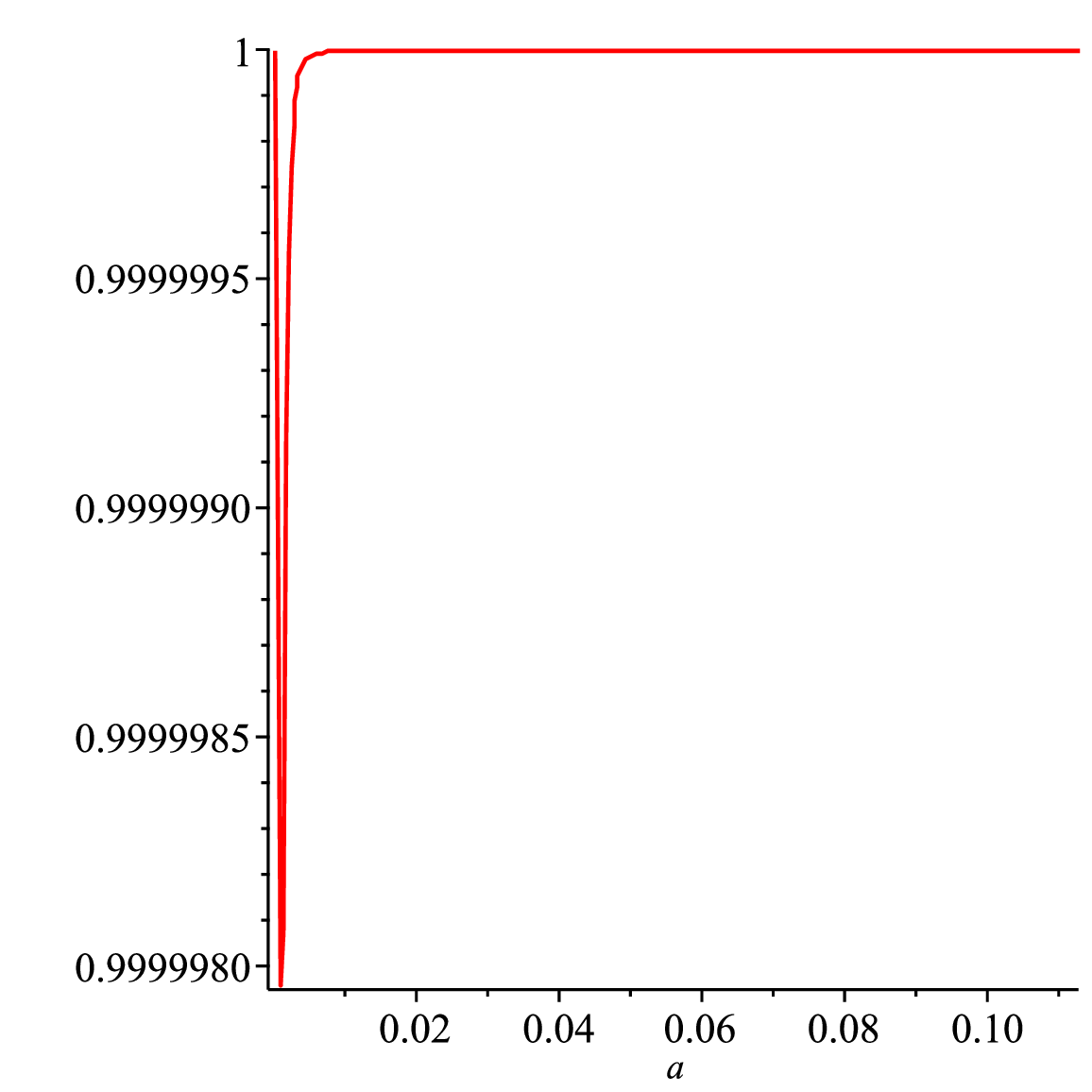}
  \caption{$c_T^2$ in terms of the scale factor}
  \label{fig:sub2}
\end{subfigure}
\caption{$c_T^2$ and $c_s^2$ versus the scale factor}
\label{fig4}
\end{figure}

\begin{figure}[H]
	\centering
	\includegraphics[scale=0.4]{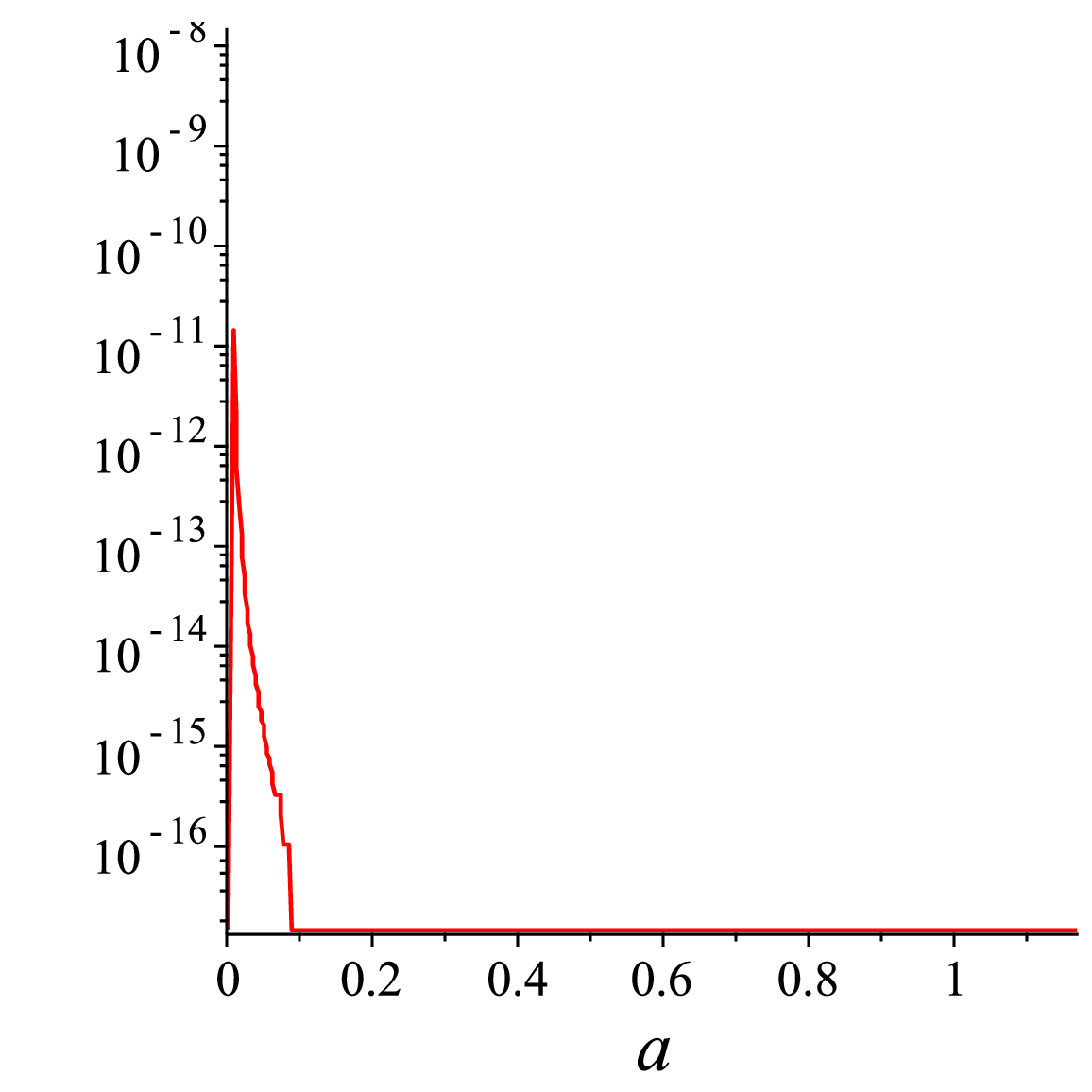}
	\caption{$1-c_T$ in terms of the scale factor}
\label{fig5}
\end{figure}

 $\hat{q}_s:=\frac{q_s}{M_P^2}$, and  $\hat{q}_T:=\frac{q_T}{M_P^2}$ are depicted in Fig.(\ref{fig3}), showing the model is free from ghosts: $\hat{q}_s>0$, and  $\hat{q}_T>0$.
We have also depicted $c_s^2>0$ as shown in Fig.(\ref{fig4}), and also $c_T^2>0$ which implies that the model has no Laplacian instability.
These figures demonstrate that the parameters closely align with their $\Lambda$CDM counterparts, $\hat{q}_T=1$, $\hat{q}_s=2$, $\hat{c}_s^2=\hat{c}_T^2=1$, except for a brief interval where the EDE becomes significant. However, as they remain positive, this deviation does not result in instability.
Finally $1-c_T$ is depicted in Fig.(\ref{fig5}),
which shows that the model respects the constraint (\ref{17}) at low redshifts.

\section{Conclusion}
We introduced a scalar-tensor model consisting of a quintessence, $\phi$, with an even potential $V(\phi)$, coupled to the Gauss-Bonnet (GB) invariant through an even function $f(\phi)$. We showed provided that $f_{,\phi \phi}(0)<0$, the quintessence is trapped in the stable fixed point $\phi=0$ for scale factors less than a critical value $a<a_c$ resulting in a standard cosmological model with an initial cosmological constant. As the universe expands, the relative dark energy density increases, and the quintessence effective mass squared, which depends on density, decreases. Provided that $V_{,\phi \phi}(0)<0$, at $a=a_c$ the quintessence becomes tachyonic and gains a nontrivial solution, causing a rapid (for a steep potential) decrease in relative initial dark energy density. This is somehow similar to scalarization where the scalar field gains a nontrivial solution when becomes tachyonic, and by its emergence, the cosmological model deviates from the standard one derived from general relativity (GR). The GWS constraint quoted in \cite{ob3}, favors a vanishingly small GB-quintessence coupling $f_{,\phi \phi}(0)$ which for a quadratic coupling is satisfied by taking $a_c$ in the radiation-dominated era. The constraint also favors a very slowly rolling quintessence at the late time, therefore we have chosen a potential that becomes eventually flat. In this way, the constructed model describes a dynamical dark energy component that becomes significant in the matter radiation equality era and then decreases rapidly and behaves as a late-time cosmological constant, which is much less than the initial cosmological constant. This model proposes a mechanism by which dark energy could have had a higher early value than what is derived from redshift calculations alone but decreased rapidly after the matter-radiation equality era. Although our model introduces an early dark energy component that can reduce the sound horizon at recombination, determining whether an early dark energy (EDE) model can resolve the Hubble tension requires a comprehensive statistical analysis incorporating all cosmological data. Additionally, EDE cosmological perturbations may be crucial in addressing the Hubble tension \cite{EDE11}.

\end{document}